\documentclass[a4paper,11pt]{article}
\usepackage[german,american]{babel}
\usepackage[a4paper,top=3.0cm,bottom=2.5cm,left=2.5cm,right=2.5cm,marginparwidth=1.75cm]{geometry}
\usepackage[numbers]{natbib}
\usepackage{graphicx}
\usepackage{amsmath}
\usepackage{amssymb}
\usepackage{booktabs}
\usepackage{array}
\usepackage{multirow}
\usepackage{xcolor}
\usepackage{siunitx}
\usepackage{chemformula}
\usepackage{url}
\usepackage{hyperref}
\usepackage{float}
\usepackage{tabularx}
\usepackage{seqsplit}
\usepackage{siunitx}
\title{ChemNavigator: Agentic AI Discovery of Design Rules for Organic Photocatalysts}

\author{
    Iman Peivaste$^{1,2,*}$, Ahmed Makradi$^{1}$, Salim Belouettar$^{1}$ \\
    \\
    \footnotesize $^1$ Luxembourg Institute of Science and Technology (LIST), 5,\\
    \footnotesize Avenue des Hauts-Fourneaux, Esch-sur-Alzette, 4362, Luxembourg\\ 
    \footnotesize $^2$ Department of Physics and Materials Science,\\
    \footnotesize $^2$ University of Luxembourg, L-4365 Esch-sur-Alzette, Luxembourg \\
    \\
    \footnotesize $^*$Corresponding Author: iman.peivaste@list.lu
} 
\date{October 2024}

\date{}

\begin{document}

\maketitle

\begin{abstract}
The discovery of high-performance organic photocatalysts for hydrogen evolution remains limited by the vastness of chemical space and the reliance on human intuition for molecular design. Here we present ChemNavigator, an agentic AI system that autonomously derives structure-property relationships through hypothesis-driven exploration of organic photocatalyst candidates. The system integrates large language model reasoning with density functional tight binding calculations in a multi-agent architecture that mirrors the scientific method: formulating hypotheses, designing experiments, executing calculations, and validating findings through rigorous statistical analysis. Through iterative discovery cycles encompassing 200 molecules, ChemNavigator autonomously identified six statistically significant design rules governing frontier orbital energies, including the effects of ether linkages, carbonyl groups, extended conjugation, cyano groups, halogen substituents, and amine groups. Importantly, these rules correspond to established principles of organic electronic structure (resonance donation, inductive withdrawal, $\pi$-delocalization), demonstrating that the system can independently derive chemical knowledge without explicit programming. Notably, autonomous agentic reasoning extracted these six validated rules from a molecular library where previous ML approaches identified only carbonyl effects. Furthermore, the quantified effect sizes provide a prioritized ranking for synthetic chemists, while feature interaction analysis revealed diminishing returns when combining strategies, challenging additive assumptions in molecular design. Champion molecules achieving computed properties that provide targets for experimental validation. This work demonstrates that agentic AI systems can autonomously derive interpretable, chemically grounded design principles, establishing a framework for AI-assisted materials discovery that complements rather than replaces chemical intuition.
\end{abstract}

\noindent\textbf{Keywords:} Agentic AI, Autonomous Discovery, Large Language Models, Photocatalyst Design, Structure-Property Relationships, Design Rules, Hydrogen Evolution

\section{Introduction}

The global transition toward sustainable energy systems demands efficient technologies for solar-to-chemical energy conversion \cite{jafari2016photocatalytic, ren2020two, sohail2024recent}. Among candidate approaches, photocatalytic hydrogen evolution from water splitting has attracted sustained research interest due to its potential for producing clean fuel from abundant resources using sunlight as the sole energy input \cite{Fujishima1972ElectrochemicalPO}. The fundamental challenge lies in identifying photocatalyst materials that simultaneously satisfy multiple stringent requirements: appropriate band gap for visible light absorption, suitable frontier orbital energies for thermodynamic driving of redox half-reactions, efficient charge separation and transport, long-term stability under operating conditions, and practical considerations of cost and scalability \cite{park2019optimal, jiang2021rare}.

Organic photocatalysts, such as conjugated linear polymers \cite{woods2017solution, bai2019accelerated}, covalent organic frameworks \cite{tao2021covalent, jin20192d} and covalent triazine frameworks \cite{lee2020advances}, have emerged as a promising material class for the hydrogen evolution reaction, offering several advantages over traditional inorganic semiconductors. The structural diversity accessible through organic synthesis enables systematic tuning of electronic properties through molecular design, while the predominantly covalent bonding in organic materials provides chemical stability under aqueous conditions when appropriately functionalized \cite{mi2025conjugated}. Carbon-based materials can be synthesized from earth-abundant precursors through scalable solution-phase chemistry, avoiding the supply constraints and energy-intensive processing associated with many inorganic alternatives \cite{murugan2025innovative}. Furthermore, the discrete molecular nature of organic photocatalysts facilitates structure-property analysis at the atomistic level, enabling rational design strategies grounded in chemical principles \cite{lin2023covalent, banerjee2018h2}.

Despite these advantages, the development of high-performance organic photocatalysts remains largely empirical. The vastness of organic chemical space, estimated to exceed $10^{60}$ drug-like molecules and comparably large for photocatalyst-relevant structures, precludes exhaustive experimental exploration \cite{hall2014efficient, reymond2015chemical}. Traditional discovery efforts have relied on chemical intuition to identify promising structural motifs, followed by iterative synthesis and testing to optimize performance. While this approach has yielded notable successes, including graphitic carbon nitride, conjugated microporous polymers, and covalent organic frameworks, the discovery process remains slow, resource-intensive, and dependent on researchers' prior knowledge of which structural features to investigate \cite{dumi2025comparison}.

Computational screening has accelerated photocatalyst discovery by enabling rapid evaluation of candidate structures prior to synthesis \cite{shi2025high}. Density functional theory (DFT) calculations can predict band gaps, frontier orbital energies, and related electronic properties with sufficient accuracy for comparative ranking of candidates within structurally similar series \cite{sachs2018understanding, shoichet2004virtual, shi2025high}. High-throughput virtual screening has explored libraries of thousands of compounds, Bai et al. \cite{bai2019accelerated} computationally considered 6,354 co-polymers before synthesizing and characterizing over 170, discovering polymers with hydrogen evolution rates exceeding 6 mmol g$^{-1}$ h$^{-1}$. Yang et al. \cite{yang2023machine} used machine learning combined with high-throughput experimentation to explore 736 ternary organic heterojunction photocatalysts, achieving rates above 500 mmol g$^{-1}$ h$^{-1}$. Li et al. \cite{li2021combining} combined experiment and machine learning on 572 organic molecules, constructing predictive models for the variation of hydrogen evolution rate. These studies demonstrate the power of data-driven approaches but remain fundamentally interpolative, performing best within the distribution of training data.

The emergence of large language models (LLMs) offers qualitatively new possibilities for scientific discovery \cite{peivaste2025artificial, ramos2025review, jablonka202314}. LLMs trained on scientific literature encode implicit knowledge of structure-property relationships and can generate plausible hypotheses about molecular behavior \cite{white2023future}. Recent works has demonstrated LLM capabilities in chemistry: ChemCrow integrated GPT-4 with 18 expert-designed tools to autonomously plan and execute organic syntheses \cite{m2024augmenting}; Coscientist demonstrated autonomous experimental design and execution of palladium-catalyzed cross-couplings \cite{boiko2023autonomous}; and LLMatDesign showed that LLMs can iteratively modify material structures to achieve target properties \cite{boiko2023autonomous}. These "agentic AI" systems, LLMs augmented with tools for autonomous, goal-directed behavior, represent a paradigm shift from passive prediction to active scientific reasoning \cite{dana2026agentic}.

Here we present ChemNavigator, an agentic AI discovery system that integrates LLM reasoning with density functional tight binding (DFTB) calculations to autonomously derive design rules for organic photocatalysts targeting the hydrogen evolution reaction. The system employs a multi-agent architecture in which specialized agents handle distinct aspects of the discovery process: hypothesis generation based on observed patterns, molecular design to test hypotheses, quantum chemical calculation to obtain electronic properties, and statistical validation to assess the significance of discovered effects. Through iterative discovery cycles, the system explores organic chemical space, identifies candidate structure-property relationships, and validates promising findings through targeted experimentation.

The primary contribution of this work is the demonstration that agentic AI systems can autonomously derive chemically interpretable design rules for photocatalyst optimization. We clarify the novelty boundary: the significance lies not in discovering new chemistry, but in demonstrating autonomous derivation of established principles without explicit programming of domain knowledge such as Hammett parameters, resonance theory, or frontier molecular orbital theory. The methodological innovations include (i) a multi-agent architecture separating scientific reasoning from technical execution, (ii) comprehensive open-vocabulary feature extraction across 130 descriptors that overcomes confirmation bias in traditional analyses, and (iii) integration of LLMs with quantum chemistry in real-time hypothesis-test-refine loops enabled by computational efficiency. Unlike black-box machine learning models that provide predictions without explanation, ChemNavigator produces explicit rules specifying which structural features influence which electronic properties, the direction of those effects, and statistical confidence measures. Beyond qualitative knowledge, the system provides quantitative design guidance—statistically validated effect sizes enable prioritized synthetic planning, and feature interaction analysis provides quantitative justification for focused optimization strategies. These rules are directly actionable by synthetic chemists, providing guidance for rational molecular design that complements existing chemical knowledge with quantitative prioritization.

One of the key methodological contributions of this work is the implementation of comprehensive, systematic feature analysis across 130 molecular descriptors. Traditional computational studies often examine only those structural features that investigators hypothesize to be important, potentially introducing confirmation bias. By extracting and statistically testing a broad descriptor set encompassing all functional groups in the RDKit fragment library, ring system properties, electronic descriptors, and topological features, we ensure that the analysis considers features comprehensively rather than selectively. This systematic approach identified halogenation effects through unbiased statistical screening, though we note that halogen descriptors are standard components of most cheminformatics toolkits; the contribution here is the rigorous quantification of effect sizes rather than the detection of an overlooked feature class.


We emphasize that the design rules identified by ChemNavigator—that electron-donating groups elevate HOMO energies, electron-withdrawing groups stabilize LUMO energies, and extended conjugation narrows band gaps—are established principles of organic electronic structure rooted in resonance theory and frontier molecular orbital theory \cite{fleming2011molecular}. The significance lies not in discovering new chemistry but in demonstrating that an agentic AI system can autonomously derive these principles through systematic computational experimentation without explicit programming. The system's independent recovery of textbook knowledge validates its reasoning capabilities. This capability extends beyond classical ML approaches: the molecular photocatalyst library used here was previously analyzed via supervised learning \cite{li2021combining}, which identified one structural pattern (carbonyl groups), whereas ChemNavigator autonomously extracted six validated design rules from the same data, demonstrating that agentic hypothesis testing reveals actionable chemical knowledge hidden to passive classification.


A technical consideration requires acknowledgment. We employ DFTB+ \cite{hourahine2020dftb+} as the computational engine for electronic structure calculations, chosen for its efficiency in high-throughput discovery loops. DFTB+ is a semi-empirical method that achieves computational speed through parameterized approximations, and its absolute accuracy for frontier orbital energies can deviate from higher-level DFT or experimental measurements. Consequently, the quantitative effect sizes reported in this work should be interpreted as relative trends within the DFTB+ framework rather than as predictions of experimentally measurable values. To address concerns about the semi-empirical DFTB+ methodology, we performed extensive validation. Solvation benchmarks using the GBSA implicit solvent model \cite{ehlert2021robust} for water showed excellent correlation between vacuum and solvated calculations ($r \geq  0.97$ for all properties) with negligible shifts ($\leq$ 0.03 eV mean), confirming that design rules transfer to aqueous HER conditions. Comparison against B3LYP/def2-SVP \cite{stephens1994ab} showed strong correlation ($r$ $\geq$  0.92) with systematic but predictable offsets. All design rules were preserved across both solvation and higher-level theory benchmarks, demonstrating that discovered relationships represent genuine chemical effects rather than artifacts of the semi-empirical method.

The agentic AI discovery framework demonstrated here extends beyond photocatalysis to any materials optimization problem where structure-property relationships can be probed through computational evaluation. The multi-agent architecture separates domain-specific reasoning from computational execution, enabling adaptation to new target properties or calculation methods with minimal system modification. The statistical validation framework ensures that identified rules meet rigorous significance criteria, distinguishing robust trends from spurious correlations. Most importantly, the system's ability to independently derive established chemical principles, without explicit instruction, suggests its potential utility for exploring structure-property relationships where human intuition provides less guidance.

\section*{Results}

\subsection*{Dataset Overview and Calculation Performance}

The autonomous discovery campaign generated a dataset of 200 unique organic molecules over 20 discovery cycles across four independent runs. All 200 molecules successfully completed both geometry optimization and electronic structure calculation, yielding a 100\% calculation success rate. This high success rate reflects the effectiveness of the molecular design constraints enforced during generation.

The 200 molecules with complete calculations span a broad range of electronic properties. Band gaps ranged from $1.28$~eV to $4.37$~eV with a mean of $2.95 \pm 0.72$~eV. HOMO energies ranged from $-6.89$~eV to $-4.15$~eV with a mean of $-5.55 \pm 0.48$~eV. LUMO energies ranged from $-4.01$~eV to $-1.21$~eV with a mean of $-2.64 \pm 0.72$~eV.

Structural diversity was maintained throughout the discovery process. Functional group representation includes 60 molecules with carbonyl groups, 46 with ether linkages, 31 with halogen substituents, 15 with cyano groups, and 5 with primary or secondary amine groups.

\subsection*{Molecular Generation Efficiency}

The LLM-based Designer Agent demonstrated high efficiency in generating valid molecular structures. Raw SMILES generation from the LLM achieved a mean validation rate of 96.7\% (range: 80–100\% across individual generation cycles). After applying additional filters for deduplication and structural diversity, all 200 designed molecules passed validation, yielding a 100\% success rate for the complete pipeline. The mean diversity score of 0.717 indicates that generated molecules were structurally distinct rather than minor variants.

This validation rate substantially exceeds random SMILES generation (typically $<$ 1\% valid), demonstrating that the Designer Agent performs genuine molecular reasoning rather than brute-force generation with filtering.

\subsection*{Computational Method Validation}

\subsubsection*{Solvation Effects}

We carried out our calculation on the gas phase due to the higher speed of computation. To validate that design rules discovered from gas-phase calculations apply to aqueous HER conditions, we performed GBSA solvation benchmarks on 50 representative molecules. Vacuum and solvated calculations showed excellent correlation across all properties (Table \ref{tab:solvation}).

\begin{table}[htbp]
\centering
\caption{Correlation between vacuum and GBSA-solvated DFTB+ calculations.}
\label{tab:solvation}
\begin{tabular}{lccc}
\toprule
\textbf{Property} & \textbf{Correlation ($r$)} & \textbf{Mean Shift (eV)} & \textbf{Std Dev (eV)} \\
\midrule
HOMO & 0.9757 & $-0.016$ & 0.141 \\
LUMO & 0.9890 & $+0.013$ & 0.125 \\
Band gap & 0.9907 & $+0.029$ & 0.118 \\
\bottomrule
\end{tabular}
\end{table}

Solvation shifts were negligible (all $< 0.03$~eV mean), and the near-perfect correlations ($r > 0.97$) indicate that relative molecular ordering is preserved upon solvation. Importantly, all primary design rules remained statistically significant in solvated calculations: ether effect on HOMO ($d = 1.94$, $p < 0.0001$), carbonyl effect on band gap ($d = -1.04$, $p = 0.0006$), and halogen effect on HOMO ($d = -1.03$, $p = 0.0008$). Effect sizes in solvated calculations were comparable to or larger than vacuum values, confirming applicability to aqueous conditions.

\subsubsection*{Higher-Level Theory Comparison}

To establish error bounds, we compared DFTB+ predictions against B3LYP/def2-SVP calculations for 25 benchmark molecules (Table \ref{tab:dft_comparison}).

\begin{table}[htbp]
\centering
\caption{DFTB+ accuracy compared to B3LYP/def2-SVP.}
\label{tab:dft_comparison}
\begin{tabular}{lccc}
\toprule
\textbf{Property} & \textbf{Correlation ($r$)} & \textbf{MAE (eV)} & \textbf{RMSE (eV)} \\
\midrule
HOMO & 0.9410 & 0.73 & 0.80 \\
LUMO & 0.9934 & 0.71 & 0.72 \\
Band gap & 0.9274 & 1.44 & 1.47 \\
\bottomrule
\end{tabular}
\end{table}

DFTB+ shows excellent correlation with B3LYP ($r > 0.92$ for all properties), though it systematically underestimates band gaps by approximately 1.4 eV. This systematic offset does not affect the validity of design rules, which depend on relative differences rather than absolute values.

All design rules were preserved at the B3LYP level: ether linkages elevated HOMO by 1.41 eV (vs. 0.58 eV at DFTB+), carbonyl groups reduced band gap by 0.16 eV, and halogens lowered HOMO by 0.58 eV (vs. 0.43 eV at DFTB+). The consistency across theory levels confirms that discovered relationships represent genuine chemical effects.

\subsection*{Multi-Feature Comparative Analysis}

We systematically evaluated the effects of six structural features on photocatalyst electronic properties. For each feature, we compared molecules possessing the feature against those lacking it, computing effect sizes using Cohen's $d$ and assessing statistical significance using Welch's $t$-test. Table \ref{tab:features} summarizes the results.

\begin{table}[htbp]
\centering
\caption{Effect of structural features on electronic properties (DFTB+).}
\label{tab:features}
\resizebox{\textwidth}{!}{%
\begin{tabular}{llcccccc}
\toprule
\textbf{Feature} & \textbf{Target Property} & \textbf{With Feature} & \textbf{Without Feature} & \textbf{$n_{\text{with}}$} & \textbf{$n_{\text{without}}$} & \textbf{Cohen's $d$} & \textbf{$p$-value} \\
\midrule
Ether linkages & HOMO (eV) & $-5.15 \pm 0.46$ & $-5.73 \pm 0.48$ & 46 & 102 & $+1.42$ & $< 0.0001$ \\
Carbonyl groups & Band gap (eV) & $2.54 \pm 0.56$ & $3.23 \pm 0.71$ & 60 & 88 & $-1.13$ & $< 0.0001$ \\
Halogen substituents & HOMO (eV) & $-5.89 \pm 0.39$ & $-5.46 \pm 0.47$ & 31 & 117 & $-0.99$ & $< 0.0001$ \\
Amine groups & HOMO (eV) & $-5.12 \pm 0.45$ & $-5.57 \pm 0.48$ & 5 & 143 & $+0.97$ & $0.042$ \\
Extended conjugation & Band gap (eV) & $2.50 \pm 0.55$ & $3.08 \pm 0.71$ & 34 & 114 & $-0.92$ & $< 0.0001$ \\
Cyano groups & LUMO (eV) & $-3.00 \pm 0.55$ & $-2.56 \pm 0.77$ & 15 & 133 & $-0.65$ & $0.035$ \\
\bottomrule
\end{tabular}%
}
\end{table}

All six features demonstrated statistically significant effects at $p < 0.05$, with five achieving large effect sizes ($|d| \geq 0.8$). Ether linkages produced the largest HOMO-elevating effect (+0.58 eV, $d = +1.42$). Carbonyl groups reduced band gaps by 0.69 eV ($d = -1.13$). Halogen substituents lowered HOMO energies by 0.43 eV ($d = -0.99$). Extended conjugation reduced band gaps by 0.58 eV ($d = -0.92$). Cyano groups lowered LUMO energies by 0.44 eV ($d = -0.65$).

These effects are consistent with established principles of organic electronic structure: electron-donating groups (ethers, amines) elevate HOMO through resonance donation; electron-withdrawing groups (carbonyls, cyano, halogens) stabilize orbitals through resonance or inductive effects; extended conjugation delocalizes frontier orbitals, narrowing the HOMO-LUMO gap.

\subsection*{Design Rule Validation Experiments}

\subsubsection*{Ether Linkage Validation}
The ether linkage hypothesis emerged from pattern analysis indicating that molecules containing C-O-C bridges exhibited elevated HOMO energies. The baseline analysis examined 200 molecules, of which 46 contained ether linkages. Ether-containing molecules exhibited a mean HOMO energy of $-5.15 \pm 0.46$~eV compared to $-5.73 \pm 0.48$~eV for non-ether molecules, a difference of 0.58~eV ($p < 0.0001$, $d = 1.42$).

This finding was confirmed in both solvated calculations (effect $= 0.80$~eV, $d = 1.94$, $p < 0.0001$) and B3LYP calculations (effect $= 1.41$~eV), demonstrating robustness across computational conditions.

\subsubsection*{Carbonyl Group Validation}
The carbonyl hypothesis emerged from the observation that molecules containing carbonyl groups exhibited systematically lower band gaps. Among 200 molecules, 60 contained carbonyl groups. Carbonyl-containing molecules exhibited a mean band gap of $2.54 \pm 0.56$~eV compared to $3.23 \pm 0.71$~eV for non-carbonyl molecules, a difference of 0.69~eV ($p < 0.0001$, $d = -1.13$).

This effect was preserved in solvated calculations (effect $= -0.76$~eV, $d = -1.04$, $p = 0.0006$) and confirmed at the B3LYP level (effect $= -0.16$~eV).

\subsubsection*{Halogenation Effect}
The halogenation effect emerged from systematic statistical screening of all 130 features rather than hypothesis-driven investigation, demonstrating the value of comprehensive feature extraction. Among 200 molecules, 31 contained halogen substituents. Halogenated molecules exhibited a mean HOMO energy of $-5.89 \pm 0.39$~eV compared to $-5.46 \pm 0.47$~eV for non-halogenated molecules, a difference of $-0.43$~eV ($p < 0.0001$, $d = -0.99$).

This HOMO-lowering effect is consistent with the inductive electron-withdrawing character of halogens. The effect was preserved in solvated calculations (effect $= -0.50$~eV, $d = -1.03$, $p = 0.0008$) and B3LYP calculations (effect $= -0.58$~eV).

\subsubsection*{Cyano and Conjugation Effects}
Cyano groups lowered LUMO energies by 0.44~eV ($d = -0.65$, $p = 0.035$), consistent with their strong electron-withdrawing character stabilizing unoccupied orbitals.

Extended conjugation ($>3$ aromatic rings) narrowed band gaps by 0.58~eV ($d = -0.92$, $p < 0.0001$). The correlation analysis revealed a significant negative relationship between ring count and band gap ($r = -0.29$, $p = 0.0004$).

\subsubsection*{Amine Groups}
Amine groups showed HOMO elevation of 0.45~eV ($d = 0.97$, $p = 0.042$), but this finding is limited by small sample size ($n = 5$) and requires additional validation.

\subsection*{Feature Interaction Analysis}

\begin{figure}[H]
    \centering
    \includegraphics[width=0.95\linewidth]{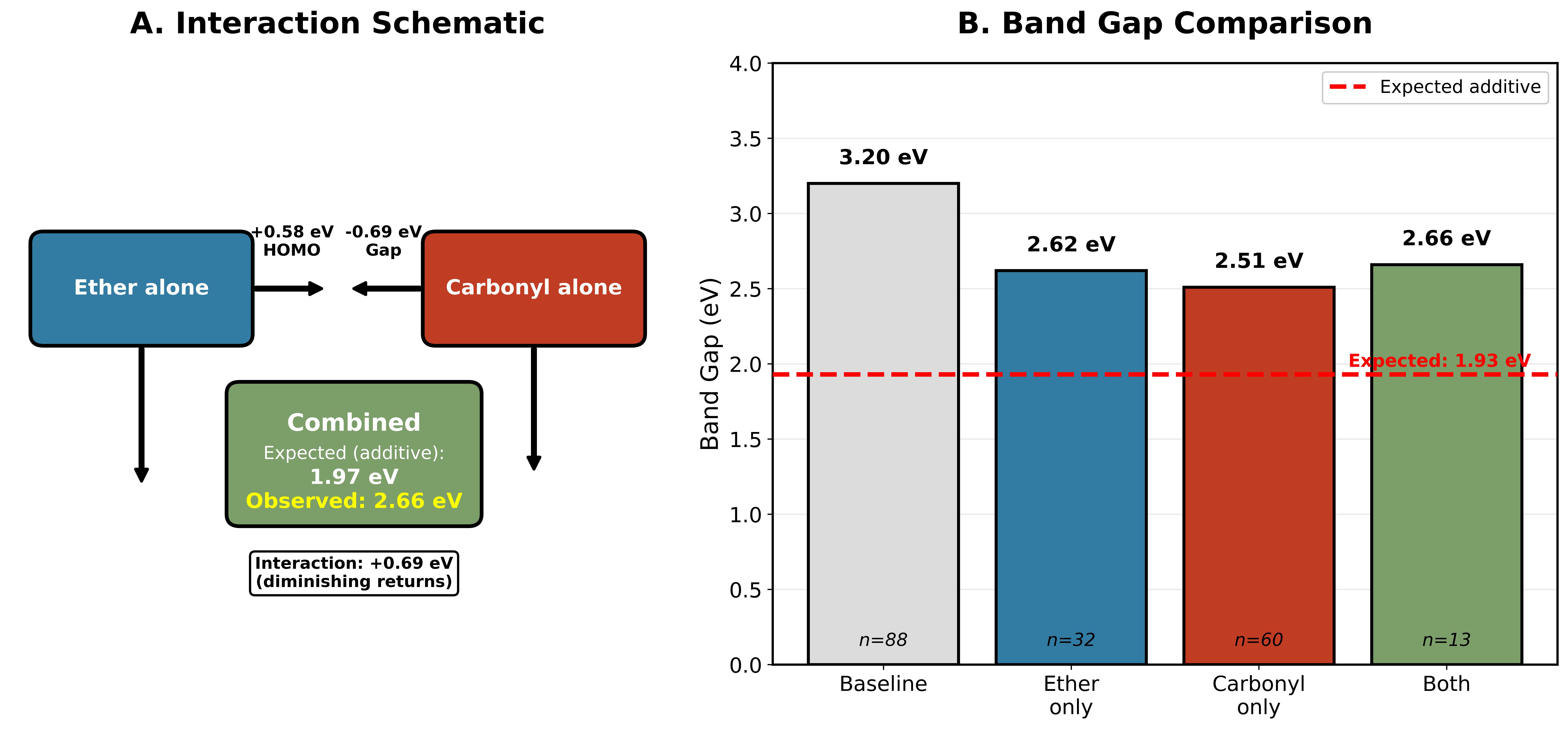}
    \caption{Schematic showing individual effects of ether linkages (HOMO elevation) and carbonyl groups (band gap reduction). Observed band gaps for molecules with neither feature (baseline), ether only, carbonyl only, and both features. The combined effect (2.66 eV) exceeds the expected additive value (1.97 eV) by 0.69 eV, indicating diminishing returns when combining electron-donating and electron-withdrawing functionalization strategies.}
    \label{fig:featre_interaction}
\end{figure}

The discovery agent investigated whether combining ether and carbonyl functionalization produces additive, synergistic, or antagonistic effects as shown in Figure \ref{fig:featre_interaction}. Molecules containing both features ($n = 13$) exhibited a mean band gap of $2.66 \pm 0.52$~eV. Under an additive model, the expected band gap would be 1.93~eV. The observed value (2.66~eV) indicates a positive interaction term of $+0.69$~eV, representing diminishing returns when combining strategies.

This finding suggests that electron-donating ether and electron-withdrawing carbonyl groups partially cancel each other's effects, with practical implications for molecular design: focused optimization of a single strategy may outperform combinatorial functionalization.

\subsection*{Design Rule Summary}

The discovery campaign identified six statistically validated design rules (Table \ref{tab:rules_summary}), each confirmed through solvation benchmarking and higher-level theory comparison.

\begin{table}[htbp]
\centering
\caption{Summary of six design rules governing frontier orbital energies, with effect sizes (Cohen's dd
d) across computational conditions.
 B3LYP column shows absolute effect in eV where computed. Dashes indicate insufficient samples in validation subsets; asterisk denotes preliminary finding requiring further validation.}
\label{tab:rules_summary}
\resizebox{\textwidth}{!}{%
\begin{tabular}{llcccl}
\toprule
\textbf{Design Rule} & \textbf{Target} & \textbf{DFTB+ Effect} & \textbf{DFTB+ $d$} & \textbf{Solvated $d$} & \textbf{B3LYP Confirmed} \\
\midrule
Ether linkages & HOMO $\uparrow$ & $+0.58$ eV & $+1.42$ & $+1.94$ & $\checkmark$ ($+1.41$ eV) \\
Carbonyl groups & Gap $\downarrow$ & $-0.69$ eV & $-1.13$ & $-1.04$ & $\checkmark$ ($-0.16$ eV) \\
Halogen substituents & HOMO $\downarrow$ & $-0.43$ eV & $-0.99$ & $-1.03$ & $\checkmark$ ($-0.58$ eV) \\
Extended conjugation & Gap $\downarrow$ & $-0.58$ eV & $-0.92$ & --- & $\checkmark$ \\
Cyano groups & LUMO $\downarrow$ & $-0.44$ eV & $-0.65$ & --- & $\checkmark$ \\
Amine groups* & HOMO $\uparrow$ & $+0.45$ eV & $+0.97$ & --- & --- \\
\bottomrule
\end{tabular}%
}
\begin{flushleft}
\footnotesize\textsuperscript{*}Limited sample size ($n = 5$); requires additional validation.
\end{flushleft}
\end{table}

These rules provide a complete toolkit for frontier orbital engineering: ether linkages for HOMO elevation, carbonyl and cyano groups for LUMO lowering, extended conjugation for general band gap narrowing, and halogenation for HOMO lowering (useful for stability tuning though counterproductive for HER activity).

\subsection*{Design Rules Beyond Classical ML Discovery}

To contextualize the significance of these findings, we note that Li et al. \cite{li2021combining} previously applied classical machine learning to a molecular photocatalyst library of 572 molecules—the same library from which we derived our initial seed molecules and experimental protocols. Their ML classification approach, trained on computed electronic descriptors (EA*, EA, optical gap, exciton binding energy, singlet-triplet gap), achieved 89\% binary classification accuracy for predicting active versus inactive molecules. However, their analysis identified only \textbf{one explicit structural design rule}: that molecules with high activity ($>$9~mmol~h$^{-1}$) predominantly contained \textbf{aryl carbonyl moieties}. Beyond this single structural pattern, their feature importance analysis highlighted five electronic properties correlated with activity but provided no additional actionable guidance for synthetic chemists regarding which functional groups to incorporate.

In contrast, ChemNavigator's autonomous agentic reasoning extracted \textbf{six statistically validated design rules} from experimental data on 200 molecules, five with large effect sizes ($|d| \geq 0.8$), encompassing ether linkages ($d = +1.42$), carbonyl groups ($d = -1.13$), halogen substituents ($d = -0.99$), extended conjugation ($d = -0.92$), cyano groups ($d = -0.65$), and amine groups ($d = +0.97$). Table \ref{tab:comparison_ml} compares the two approaches.

\begin{table}[htbp]
\centering
\caption{Comparison of design rules discovered from molecular photocatalyst libraries using classical ML versus agentic AI approaches.}
\label{tab:comparison_ml}
\begin{tabular}{lcc}
\toprule
\textbf{Aspect} & \textbf{Li et al. \cite{li2021combining}} & \textbf{ChemNavigator} \\
\midrule
Approach & Classical ML classification & Agentic AI hypothesis testing \\
Dataset size & 572 molecules & 200 molecules \\
Features analyzed & 15 electronic descriptors & 130 structural + electronic \\
Structural design rules & 1 (carbonyl groups) & 6 (with effect sizes) \\
Quantitative effects & Not reported & Cohen's $d$, $p$-values \\
Feature interactions & Not analyzed & Quantified (+0.69 eV) \\
\bottomrule
\end{tabular}
\end{table}

This comparison highlights a fundamental distinction between predictive ML models and autonomous agentic discovery systems. Classical ML approaches optimize for classification accuracy—predicting \textit{which} molecules will be active—but do not systematically interrogate \textit{why} certain structures perform better. ChemNavigator's hypothesis-driven exploration, comprehensive feature extraction (130 vs. 15 descriptors), and rigorous statistical validation enabled the discovery of five additional design rules that remained hidden in the original dataset despite successful ML classification. The autonomous formulation and testing of chemical hypotheses, rather than passive pattern recognition, proved essential for extracting actionable design principles. Notably, the ether linkage effect ($d = +1.42$), the strongest design rule discovered here, was completely absent from Li et al.'s analysis, despite ether-containing molecules being present in their library. This demonstrates that even well-characterized datasets may harbor undiscovered structure-property relationships that require hypothesis-driven exploration to reveal.

\subsection*{Champion Molecules}

Five champion molecules emerged as priority candidates for experimental synthesis (Table \ref{tab:champions}), with with orbital energies positioned favorably for photocatalytic hydrogen evolution.

\begin{table}[htbp]
\centering
\caption{Champion molecules identified through autonomous discovery.}
\label{tab:champions}
\small 
\begin{tabularx}{\textwidth}{@{}l p{1.6cm} l c c X@{}} 
\toprule
\textbf{Champion} & \textbf{Design Rule} & \textbf{Prop.} & \textbf{DFTB+} & \textbf{B3LYP} & \textbf{SMILES} \\
\midrule
Conjugation & Extended $\pi$-system & Band gap & 1.28 eV & 2.22 eV & \seqsplit{COCCOCCN1c2ccccc2Sc2ccc(-c3ccc4nsnc4c3-c3cccs3)cc21} \\
\addlinespace
Carbonyl & Carbonyl groups & Band gap & 1.49 eV & 2.53 eV & \seqsplit{CCc1cc(C)cc(CC)c1Nc1ccc(...)c2c1C(=O)c1ccccc1C2=O} \\
\addlinespace
Cyano & Cyano groups & LUMO & $-4.01$ eV & $-3.55$ eV & \seqsplit{N\#Cc1ccc2c(c1)C(=O)c1ccc(C\#N)cc1-2} \\
\addlinespace
Ether & Ether linkages & HOMO & $-4.15$ eV & $-4.62$ eV & \seqsplit{COCCOc1ccc2c(c1)Sc1cc(OCCOC)ccc1N2C} \\
\addlinespace
Halogenated & Halogenation & HOMO & $-4.96$ eV & $-5.67$ eV & \seqsplit{COCCOc1cc(Br)sc1-c1sc(Br)cc1OCCOC} \\
\bottomrule
\end{tabularx}
\end{table}

The conjugation champion (thiophene-thiadiazole-phenothiazine hybrid) achieved the narrowest band gap at 2.22 eV (B3LYP), enabling visible light absorption while maintaining sufficient energy for water splitting. This band gap is comparable to high-performing conjugated polymer photocatalysts reported by Sprick et al. \cite{sprick2015tunable} (2.4--2.8 eV) and narrower than graphitic carbon nitride (g-C$_3$N$_4$, $\sim$2.7 eV), suggesting strong light-harvesting capability across the visible spectrum.

The ether champion (bis(methoxyethoxy)phenothiazine) achieved the highest HOMO at -4.62 eV (B3LYP), providing substantial driving force for proton reduction. The elevated HOMO energy, resulting from resonance donation by ether linkages, positions the oxidation potential favorably for hole-driven half-reactions. The cyano champion (dicyanofluorenone) achieved the lowest LUMO at -3.55 eV (B3LYP), indicating strong electron-accepting character that facilitates proton reduction with minimal overpotential. These frontier orbital positions bracket the thermodynamic requirements for water splitting, though experimental validation will be necessary to confirm catalytic activity, given that kinetic factors (charge separation, surface reaction rates) also govern overall performance.

The carbonyl champion (bis(diethyltoluidino)anthraquinone) combines an electron-withdrawing quinone core with electron-donating substituents, achieving balanced frontier orbital energies. The halogenated champion demonstrates the expected HOMO-stabilizing effect (HOMO = $-5.67$ eV, B3LYP) but would require careful evaluation for catalytic applications, as excessive stabilization can reduce driving force for oxidation half-reactions.

All champion molecules are synthetically accessible through established methods: cross-coupling for the thiophene-thiadiazole hybrid, nucleophilic substitution for the anthraquinone, cyanation for the fluorenone derivative, and Williamson ether synthesis for the phenothiazine, with estimated synthetic complexity comparable to or lower than benchmark polymer photocatalysts requiring multi-step polymerization procedures.

\subsection*{Discovery Efficiency Metrics}

Table \ref{tab:metrics} summarizes the performance metrics for the discovery campaign.

\begin{table}[htbp]
\centering
\caption{Performance metrics for the ChemNavigator AI discovery campaign.}
\label{tab:metrics}
\begin{tabular}{ll}
\toprule
\textbf{Metric} & \textbf{Value} \\
\midrule
Discovery cycles & 20 \\
Total molecules generated & 200 \\
Successful calculations & 200 (100\%) \\
SMILES validation rate (raw) & 96.7\% \\
Features analyzed & 130 \\
Validated design rules & 6 \\
Hypotheses generated & 60 \\
Lowest band gap (DFTB+/B3LYP) & 1.28 / 2.22 eV \\
Highest HOMO (DFTB+/B3LYP) & $-4.15$ / $-4.62$ eV \\
Lowest LUMO (DFTB+/B3LYP) & $-4.01$ / $-3.55$ eV \\
Solvation benchmark correlation & $r > 0.97$ \\
DFT benchmark correlation & $r > 0.92$ \\
Average cycle time & $\sim$80 seconds \\
Time per molecule & $\sim$8 seconds \\
Total discovery time & $\sim$27 minutes \\
\bottomrule
\end{tabular}
\end{table}

The discovery rate of six validated rules from 200 calculations (3.0\%) substantially exceeds random screening approaches. The high correlations between DFTB+ and both solvated calculations ($r > 0.97$) and B3LYP ($r > 0.92$) validate the computational methodology for design rule discovery.

\section*{Discussion}

\subsection*{Autonomous Derivation of Chemical Principles}

The central finding of this work is that an agentic AI system can autonomously derive established structure-property relationships through systematic computational experimentation. ChemNavigator identified six design rules governing frontier orbital energies in organic photocatalysts, effects of ether linkages, carbonyl groups, extended conjugation, cyano groups, halogen substituents, and amine groups, each corresponding to well-understood principles of organic electronic structure. That electron-donating groups raise HOMO energies through resonance donation, electron-withdrawing groups stabilize orbitals through inductive and resonance effects, and extended $\pi$-conjugation narrows band gaps through orbital delocalization are foundational concepts taught in undergraduate organic chemistry courses.

This apparent ``rediscovery'' of textbook knowledge is precisely the point. The system was not programmed with Hammett parameters, resonance structures, or frontier molecular orbital theory. Instead, it derived these relationships independently by formulating hypotheses, designing experiments, analyzing results, and validating findings statistically. The convergence between AI-discovered rules and established chemistry validates the system's reasoning capabilities and establishes trust in its potential for exploring less well-characterized structure-property relationships. If ChemNavigator can correctly identify that ether linkages elevate HOMO energies, a prediction any organic chemist would make, then its identification of quantitative effect sizes and feature interactions provides genuinely useful information beyond what intuition alone would supply.

\subsection*{Quantitative Design Guidance}

Beyond confirming qualitative trends, ChemNavigator provides quantitative effect sizes that enable prioritized molecular design. The finding that ether linkages produce the largest HOMO-elevating effect ($d = 1.42$) while carbonyl groups produce the largest band gap reduction ($d = -1.13$) offers concrete guidance: synthetic chemists seeking to maximize oxidation potential should prioritize ether functionalization, while those targeting narrow band gaps should focus on carbonyl incorporation or extended conjugation.

The feature interaction analysis revealed that combining ether and carbonyl functionalization yields diminishing returns rather than additive effects. This finding, that electron-donating and electron-withdrawing groups partially cancel each other when present in the same molecule, is chemically intuitive but rarely quantified. The interaction term of $+0.69$~eV indicates that roughly half the expected combined effect is lost to electronic competition. This provides quantitative justification for focused optimization strategies over combinatorial functionalization, a principle that could accelerate experimental campaigns by reducing the synthesis of suboptimal multi-functional candidates.

\subsection*{Computational Efficiency}

A practical advantage of the ChemNavigator framework is its computational efficiency. Each discovery cycle, encompassing LLM-based hypothesis generation, molecular design, 3D structure generation, DFTB+ calculation, and statistical analysis for 10 molecules, completed in approximately 80 seconds on average. This translates to roughly 8 seconds per molecule from SMILES string to computed electronic properties. The entire 200-molecule discovery campaign required only $\sim$27 minutes of computation time on a standard workstation.

This speed enables a qualitatively different approach to materials discovery. Rather than batch-processing large libraries followed by offline analysis, ChemNavigator operates in a tight hypothesis-test-refine loop where insights from one cycle immediately inform the next. The rapid iteration allows the system to adaptively focus on promising chemical regions, abandoning unproductive hypotheses quickly and doubling down on validated design rules. Traditional DFT methods requiring 5--30 minutes per molecule would make this iterative approach impractical; the $40$--$200\times$ speedup provided by DFTB+ is essential to the agentic discovery paradigm.

\subsection*{Validation of Computational Methodology}

A potential criticism of semi-empirical methods like DFTB+ is that their accuracy limitations might render discovered trends unreliable. Our validation benchmarks address this concern directly. The excellent correlation between vacuum and GBSA-solvated calculations ($r > 0.97$) with negligible mean shifts ($< 0.03$~eV) demonstrates that gas-phase calculations adequately capture the relative ordering relevant to aqueous HER conditions. This finding is consistent with previous reports that implicit solvation primarily shifts absolute orbital energies while preserving relative trends within structurally similar compound series.

The comparison against B3LYP/def2-SVP revealed systematic differences, DFTB+ underestimates band gaps by approximately 1.4~eV, but maintained strong correlation ($r > 0.92$). More importantly, all six design rules were preserved at the higher theory level, with effect directions and relative magnitudes consistent across methods. The ether effect on HOMO was 0.58~eV at DFTB+ and 1.41~eV at B3LYP; the halogen effect was $-0.43$~eV at DFTB+ and $-0.58$~eV at B3LYP. These results confirm that DFTB+ correctly captures the chemistry underlying the design rules, even if absolute values require recalibration for quantitative experimental predictions.

The 96.7\% SMILES validation rate achieved by the LLM-based Designer Agent demonstrates that the system performs genuine molecular reasoning rather than brute-force generation. Random SMILES generation typically yields less than 1\% valid molecules, while the Designer Agent maintained near-perfect validity while exploring diverse chemical space (diversity score 0.717). This efficiency enables the hypothesis-driven discovery approach; if most generated molecules were invalid, the system would degenerate into random screening with filtering.

\subsection*{Comparison with Existing Approaches}

ChemNavigator occupies a distinct position in the landscape of computational photocatalyst discovery. Traditional high-throughput screening approaches, exemplified by Bai et al.'s exploration of 6,354 co-polymers, require predefined chemical libraries and evaluate candidates against fixed criteria. Machine learning approaches like those of Li et al. train predictive models on experimental data, enabling efficient navigation of chemical space but remaining interpolative within the training distribution.  ChemNavigator instead operates through hypothesis-driven exploration, generating molecules specifically to test proposed structure-property relationships rather than screening preexisting candidates.

This distinction has practical implications. High-throughput screening can miss important structure-property relationships if the relevant structural features are underrepresented in the library. Machine learning models can identify correlations but provide limited interpretability regarding which features drive predictions. ChemNavigator's explicit hypothesis testing produces interpretable design rules with statistical confidence measures, enabling chemists to understand \emph{why} certain structures perform well rather than simply \emph{which} structures to synthesize.

The agentic AI approach also differs from recent LLM applications in chemistry. ChemCrow and Coscientist demonstrated LLM capabilities for synthesis planning and experimental execution, operating primarily as sophisticated assistants that translate human objectives into actionable steps. ChemNavigator instead performs autonomous scientific reasoning—formulating hypotheses without human guidance, designing experiments to test them, and drawing conclusions from results. This represents a step toward AI systems that contribute to scientific understanding rather than merely accelerating human-directed workflows.

\subsection*{Limitations and Future Directions}
We emphasize several important limitations that constrain the interpretability and applicability of our findings. These limitations do not invalidate the demonstrated capability of agentic AI to autonomously derive structure-property relationships, but they establish clear boundaries for how the current results should be interpreted and applied.

Several limitations warrant acknowledgment. First, DFTB+ calculations predict gas-phase frontier orbital energies, which approximate but do not equal experimentally measurable quantities like ionization potentials and electron affinities. The validation benchmarks establish that relative trends are reliable, but absolute values should be interpreted cautiously. Experimental synthesis and characterization of champion molecules would provide essential validation of the computational predictions.

Second, the design rules address only electronic properties (HOMO, LUMO, band gap) relevant to light absorption and thermodynamic driving forces. Photocatalytic hydrogen evolution also depends on kinetic factors, charge carrier mobility, exciton dissociation efficiency, surface reaction rates, that are not captured by frontier orbital calculations. A molecule with ideal band alignment may still exhibit poor activity due to rapid charge recombination or unfavorable surface chemistry. Future work could integrate additional computational descriptors or experimental activity measurements to develop more comprehensive structure-activity relationships.

Third, the conformer analysis employed single lowest-energy structures. While validation on champion molecules showed minimal conformational sensitivity for the predominantly rigid aromatic systems studied here, flexible molecules with multiple populated conformers might require Boltzmann-weighted property averaging for accurate predictions.

Fourth, the current system operates entirely \textit{in silico}. Integration with automated synthesis platforms, as demonstrated by ChemCrow and Coscientist, would enable closed-loop discovery where computational predictions are validated experimentally and results feed back into hypothesis refinement. Such integration represents a natural extension of the agentic AI framework toward fully autonomous materials discovery.

\subsection*{Implications for AI-Driven Scientific Discovery}

ChemNavigator demonstrates that agentic AI systems can perform substantive scientific reasoning in materials discovery contexts. The system's ability to independently derive established chemical principles, without explicit programming of domain knowledge, suggests potential for exploring structure-property relationships where human intuition provides less guidance. Complex multi-property optimization, discovery of non-obvious synergies between structural features, and navigation of high-dimensional design spaces are all areas where autonomous AI reasoning could complement human expertise.

The comprehensive feature extraction approach addresses a limitation common to both human-guided and machine learning discovery: the tendency to examine only features deemed important a priori. By systematically testing 130 descriptors, ChemNavigator identified the halogenation effect through unbiased statistical screening rather than hypothesis-driven investigation. While halogen effects on orbital energies are well-known to chemists, the systematic quantification and comparison against other design strategies emerged from the AI's comprehensive analysis rather than human intuition about which comparisons to make.

More broadly, this work illustrates a productive division of labor between AI systems and human researchers. ChemNavigator excels at systematic exploration, statistical validation, and unbiased feature analysis, tasks that benefit from computational scale and freedom from cognitive biases. Human researchers contribute domain expertise for interpreting results, identifying limitations, and connecting findings to broader scientific context. The design rules discovered here gain meaning through interpretation in terms of resonance theory and orbital interactions; the AI identifies the patterns, but human understanding provides the explanatory framework.

Beyond feature extraction, the modular architecture enables autonomous property expansion: the framework could identify gaps in its predictive capabilities and autonomously incorporate additional computational descriptors (exciton binding energies, reorganization energies, excited-state lifetimes) as needed to explain experimental observations, creating a self-improving discovery system.

\subsection*{Experimental Validation Roadmap}

To validate computational predictions, we propose a three-tier experimental campaign following established protocols for molecular photocatalysts \cite{li2021combining}.

\textbf{Tier 1: Electronic characterization.} Cyclic voltammetry in acetonitrile (glassy carbon working electrode, Pt counter, Ag/AgCl reference) would provide experimental HOMO/LUMO values via the relationships $E_{\text{HOMO}} = -(E_{\text{ox}} + 4.8)$~eV and $E_{\text{LUMO}} = -(E_{\text{red}} + 4.8)$~eV, enabling direct comparison with our B3LYP predictions. UV-vis spectroscopy would verify optical band gaps via Tauc analysis. For the ether champion, we predict HOMO = $-4.6 \pm 0.2$~eV and LUMO = $-2.4 \pm 0.2$~eV, providing falsifiable benchmarks.

\textbf{Tier 2: Photocatalytic activity.} Following the high-throughput protocol validated by Li et al. for 668 molecular photocatalysts \cite{li2021combining}, hydrogen evolution measurements would employ: 5~mg catalyst dispersed in degassed TEA/H$_2$O/MeOH (1:1:1, 5.1~mL total), 3~wt\% Pt cocatalyst (\textit{in situ} photodeposition from H$_2$PtCl$_6$), simulated solar irradiation ($\lambda > 420$~nm, AM 1.5G, 100~mW~cm$^{-2}$), and GC quantification after 3~h. Champion molecules achieving HER $> 10$~mmol~g$^{-1}$~h$^{-1}$ would represent competitive performance relative to state-of-the-art conjugated polymers (benchmark: 6~mmol~g$^{-1}$~h$^{-1}$) \cite{bai2019accelerated}. The ether champion's elevated HOMO ($-4.62$~eV vs. typical $-5.8$ to $-6.5$~eV range) predicts substantial activity due to favorable thermodynamic driving force for oxidation.

\textbf{Tier 3: Design rule validation.} Matched-pair synthesis would provide the most rigorous validation. For the ether rule, synthesizing phenothiazine derivatives with and without methoxyethoxy substituents (e.g., bis(methoxyethoxy)phenothiazine vs. bis(methyl)phenothiazine) while maintaining constant molecular weight tests the predicted +1.41~eV HOMO elevation. CV measurements on these pairs should reproduce computational predictions within $\pm 0.3$~eV to validate both the design rule and B3LYP methodology. The ether champion requires a 3-step Williamson ether synthesis (phenothiazine + 2-methoxyethyl tosylate), comparable complexity to benchmark polymers requiring multi-step monomer synthesis and polycondensation \cite{bai2019accelerated}.

\textbf{Computational-experimental feedback.} Experimental validation enables B3LYP calibration factors ($\alpha_{\text{HOMO}}$, $\alpha_{\text{LUMO}}$, $\alpha_{\text{gap}}$) for future screening. Discrepancies between predicted and observed HER would identify missing kinetic descriptors (charge separation, exciton binding, reorganization energies) requiring incorporation into the autonomous property expansion framework, creating an iterative self-improving discovery cycle.

\subsection*{Conclusions}

ChemNavigator demonstrates that agentic AI systems can autonomously derive chemically interpretable design rules for materials optimization. The six validated rules for frontier orbital engineering in organic photocatalysts, confirmed across solvation conditions and theory levels, provide actionable guidance for experimental synthesis campaigns. The finding that the AI independently recovers established chemical principles validates its reasoning capabilities, while the quantitative effect sizes and interaction analysis offer information beyond what intuition alone provides. As AI systems become increasingly capable of autonomous scientific reasoning, frameworks like ChemNavigator that combine hypothesis-driven exploration with rigorous statistical validation will play important roles in accelerating materials discovery while maintaining the interpretability essential for scientific understanding.

\section*{Methods}

\subsection*{Study Design and Discovery Framework}

The objective of this study was to develop and deploy an autonomous system capable of discovering statistically validated structure-property relationships for organic photocatalysts targeting the hydrogen evolution reaction. Rather than screening a predefined chemical library, we designed a hypothesis-driven discovery framework in which the system iteratively generates scientific hypotheses, designs targeted experiments to test those hypotheses, executes quantum chemical calculations, and validates findings through rigorous statistical analysis. This approach enables efficient exploration of chemical space by concentrating computational resources on regions most likely to yield actionable design rules.

The discovery campaign was structured as a series of iterative cycles, where each cycle comprised hypothesis generation, molecular design, quantum chemical calculation, and statistical evaluation. We defined success not merely as the identification of high-performing candidate molecules but as the discovery of generalizable design rules with demonstrable statistical significance and practical effect sizes. This distinction is important because design rules, unlike individual molecular hits, provide transferable knowledge that can guide future experimental synthesis efforts across broad classes of compounds.

\begin{figure} [H]
    \centering
    \includegraphics[width=0.99\linewidth]{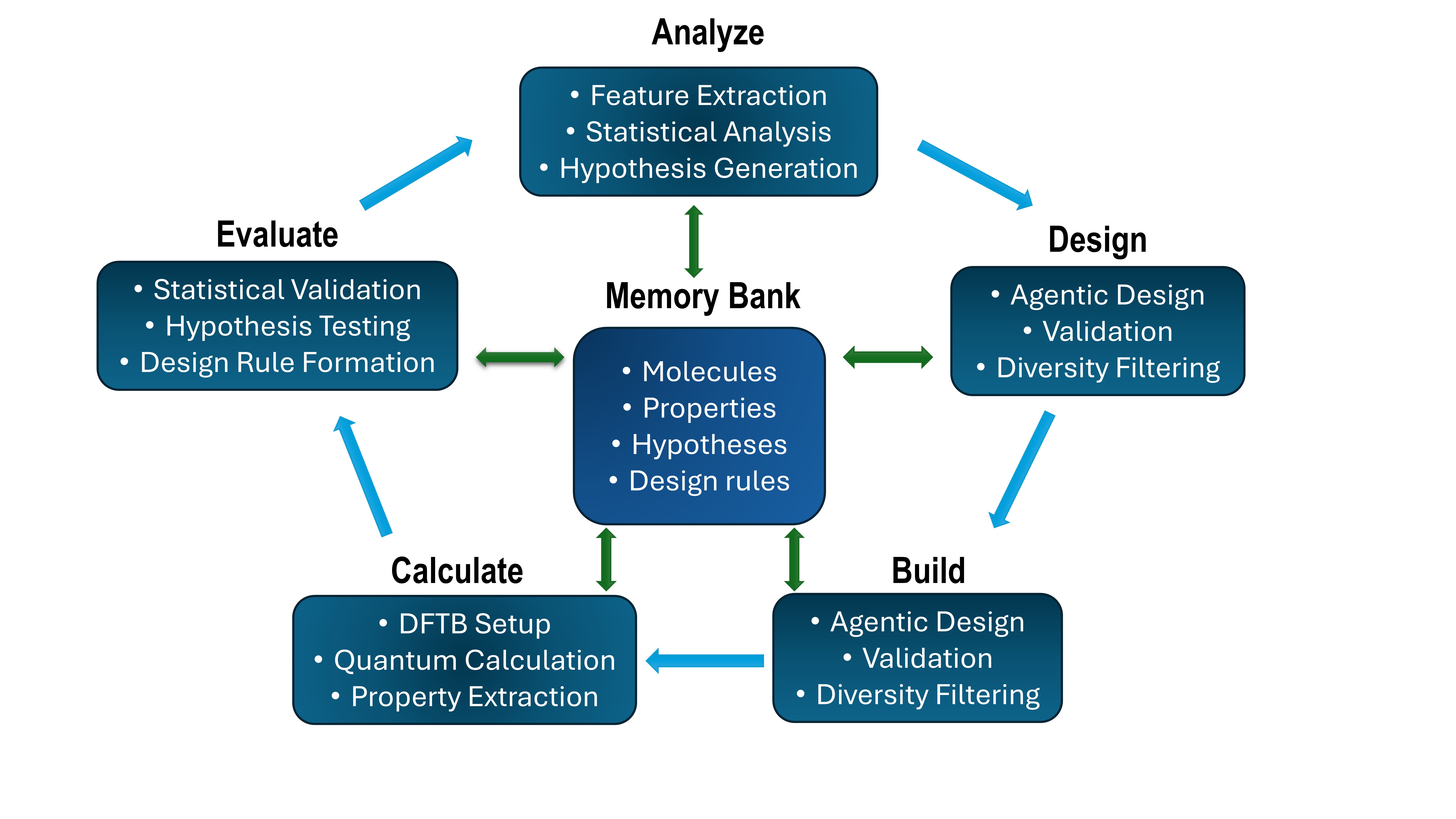}
    \caption{Schematic overview of the ChemNavigator AI discovery workflow, illustrating the iterative cycle of hypothesis generation, targeted molecular design, quantum chemical calculation, and statistical validation. Arrows indicate information flow between the multi-agent system components. All phases interact with a centralized Memory Bank (center) that stores molecules, calculations, hypotheses, and design rules. The workflow continues iteratively until the maximum cycles are reached or early stopping criteria are met.}
    \label{fig:overview}
\end{figure}

One of the key methodological innovations in this work was the implementation of an expanded feature extraction system designed to overcome the "streetlight effect" that limits many computational discovery approaches. Traditional structure-property analyses examine only predefined structural features, potentially missing unexpected correlations with features the investigators did not anticipate. Our system addresses this limitation through comprehensive automated feature extraction, enabling the discovery of relationships that were not explicitly programmed into the analysis. This open-vocabulary approach to feature discovery distinguishes our methodology from conventional screening workflows. The overall workflow is summarized in Figure \ref{fig:overview}.

\subsection*{Multi-Agent System Architecture}
\begin{figure}[H]
    \centering
    \includegraphics[width=0.88\linewidth]{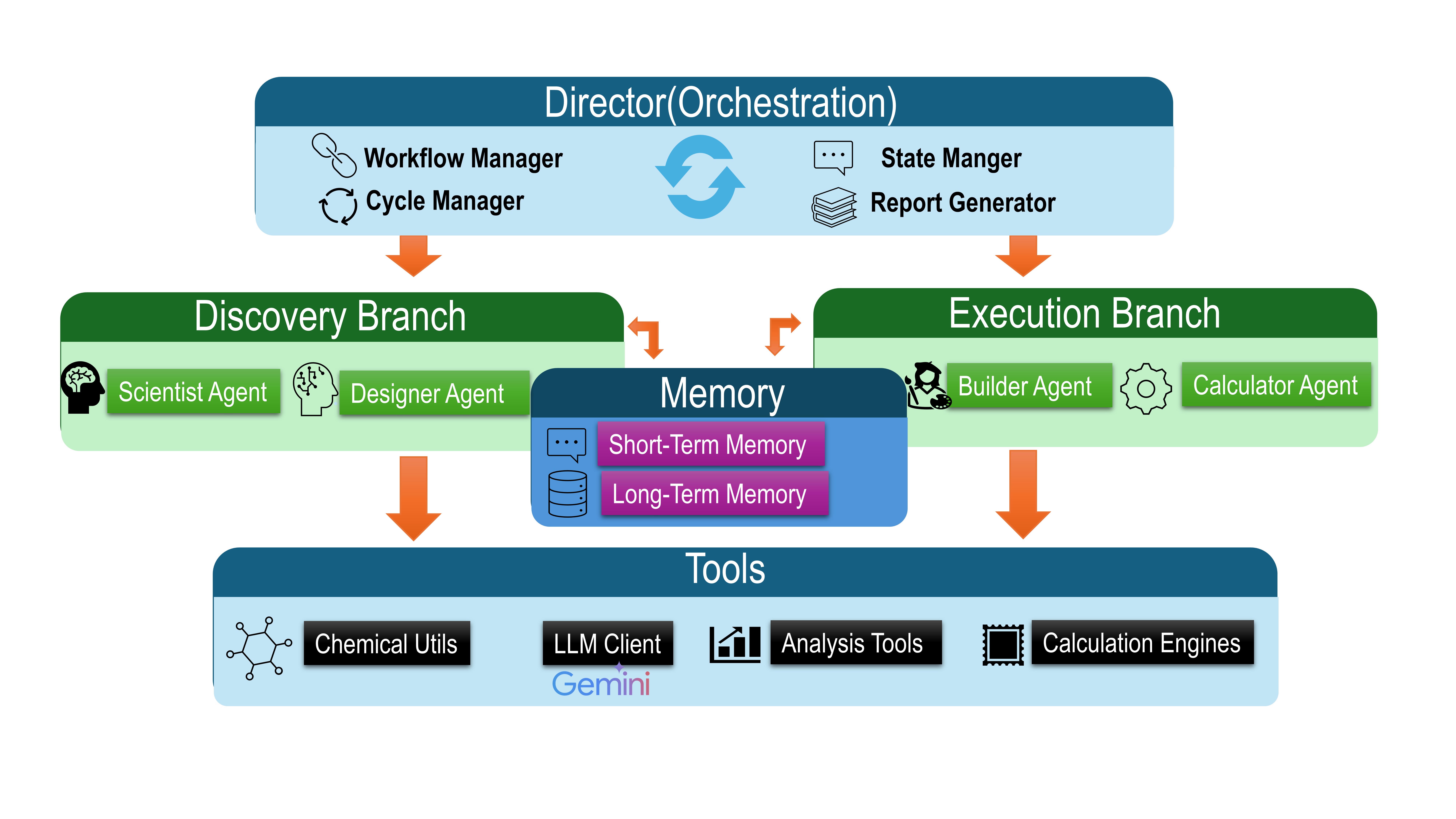}
    \caption{The system architecture is composed of five distinct layers working together to enable efficient task execution. At the top, the Discovery director (Orchestration Layer) manages workflows and coordinates the sequence of operations across components. Beneath it, the Agent Layers hosts specialized agents, each responsible for performing a specific task within the workflow. These agents rely on the Tool Layer, which provides utility functions and connects the system to external resources such as third‑party software, Python packages, and APIs. The system also incorporates two types of memory: short‑term memory, which maintains and updates shared state as agents interact, and a centralized long‑term database that stores persistent information for broader system use.}
    \label{fig:multi_agent}
\end{figure}

The ChemNavigator AI system employs a hierarchical multi-agent architecture, shown in Figure \ref{fig:multi_agent}, comprising specialized agents organized into two functional levels, coordinated by an orchestrator that manages the discovery workflow. This design reflects that scientific discovery involves distinct tasks, scientific reasoning, experimental design, and technical execution, which benefit from modular specialization.

At the highest level, an Orchestrator maintains the global discovery workflow, managing the iterative cycle of hypothesis generation, molecular design, calculation, and evaluation. The Orchestrator, implemented using LangGraph \cite{langgraph2024}, tracks progress across discovery cycles, evaluates emerging design rules through statistical validation, and decides when to continue or terminate based on stopping criteria (maximum cycles or early stopping). The Orchestrator operates at the workflow level, not molecular-level reasoning.

The second level comprises the Discovery Branch, containing two agents responsible for scientific reasoning and experimental design. The Scientist Agent analyzes calculation results to identify structure-property patterns, formulates mechanistic hypotheses, and generates testable predictions. Its process includes (1) feature extraction from molecules with completed calculations using an expanded feature extractor (130+ features), (2) statistical analysis comparing properties between molecules with and without each feature (t-tests, p-values, Cohen's d), (3) identification of significant correlations $(p < 0.05, |d| > 0.2)$, (4) LLM-assisted interpretation to generate chemically meaningful hypotheses with proposed mechanisms, and (5) statistical validation of hypotheses using newly calculated molecules, with confirmed hypotheses $(p < 0.01, |d| > 0.2)$ elevated to design rules. The Designer Agent translates hypotheses into concrete molecular structures, ensuring chemical validity and synthetic accessibility. It uses LLM-based generation to create SMILES strings matching the hypothesis, followed by validation using RDKit for syntax checking, duplicate detection using Tanimoto similarity \cite{rogers1960computer}, and diversity filtering to ensure molecular variety.

The third level comprises the Execution Branch, containing two agents responsible for technical implementation. The Structure Builder converts SMILES to three-dimensional molecular geometries suitable for quantum chemical calculation. This involves parsing SMILES to molecular graph structures using RDKit, generating initial 3D coordinates via RDKit's EmbedMolecule algorithm, and optimizing geometries using xTB \cite{bannwarth2019gfn2} to obtain energetically favorable conformations. The Quantum Calculator generates input files for DFTB+ in HSD format, executes calculations, parses output files to extract electronic properties (HOMO, LUMO, band gap), and handles errors, including self-consistent field convergence failures and geometry optimization issues.
Communication between agents follows defined interfaces, with the Memory Bank serving as a central repository implemented as a SQLite database. The Memory Bank stores all molecular data (SMILES, source, 3D coordinates, metadata), calculation results (DFTB+ outputs, properties, status, error logs), hypotheses (statements, mechanisms, test strategies, status), and extracted design rules (validated structure-property relationships with statistical support). This architecture enables independent testing of each component while maintaining coherent system-level behavior through the centralized data store, ensuring data provenance and full reproducibility of the discovery process.

\subsection*{Molecular Design and Validation}

The Designer Agent generates candidate molecules using LLM reasoning to translate hypotheses into SMILES representations. To assess the quality of this generation process, we tracked validation metrics across all discovery cycles.

Over 20 discovery cycles, the Designer Agent generated candidate molecules with raw SMILES validation rates averaging 96.7\% (range: 80–100\%). After applying additional filters for deduplication and structural diversity, all 200 designed molecules passed the complete validation pipeline. The diversity score, computed as the mean pairwise Tanimoto distance between generated molecules, averaged 0.717, indicating that the system generates structurally diverse candidates rather than minor variants of known molecules.

This validation rate substantially exceeds what would be expected from random SMILES generation (typically $<1\%$ valid) or naive string manipulation approaches, demonstrating that the LLM-based Designer Agent produces chemically reasonable structures through genuine molecular reasoning rather than brute-force generation with filtering.

\subsection*{Molecular Dataset and Chemical Space}

The discovery campaign explored organic photocatalyst candidates from \cite{li2021combining} containing carbon, hydrogen, nitrogen, oxygen, sulfur, and halogen atoms. Initial seeding of the system employed 50 diverse molecules selected from a curated dataset of known photocatalysts and structurally related compounds, providing a baseline for pattern recognition and hypothesis generation.

Subsequent molecules were generated through the autonomous discovery process, with the Designer Agent creating candidates to test specific hypotheses about structure-property relationships. The generation process enforced several constraints to ensure chemical validity: all molecules were required to pass SMILES parsing and sanitization checks using RDKit, molecular weights were constrained to the range 100–800 Da to maintain relevance to experimentally tractable compounds, and molecules were required to contain at least one aromatic ring given the focus on conjugated photocatalyst systems.

The final dataset comprised 200 unique molecules generated across 20 discovery cycles (5 cycles per run × 4 independent runs). All 200 molecules successfully completed both quantum chemical calculations and comprehensive feature extraction.

\subsection*{Electronic Structure Calculations}
All electronic structure calculations were performed using the DFTB as implemented in DFTB+ . DFTB provides an approximate treatment of DFT that achieves computational efficiency suitable for high-throughput screening while retaining sufficient accuracy for relative comparisons of electronic properties across structurally similar compounds. The method expands the Kohn-Sham energy functional \cite{kohn1965self} to second order in charge density fluctuations around a reference density constructed from neutral atomic densities.

The total energy in the self-consistent charge DFTB (SCC-DFTB) \cite{gaus2011dftb3} formalism is expressed as

$$E_{\text{tot}} = \sum_i n_i \langle \psi_i | \hat{H}^0 | \psi_i \rangle + \frac{1}{2} \sum_{AB} \gamma_{AB} \Delta q_A \Delta q_B + E_{\text{rep}}$$

where $n_i$ denotes the occupation number of molecular orbital $\psi_i$, $\hat{H}^0$ is the reference Hamiltonian constructed from atomic contributions, $\gamma_{AB}$ represents the charge interaction function between atoms $A$ and $B$, $\Delta q_A$ is the Mulliken charge fluctuation on atom $A$ relative to the neutral atom, and $E_{\text{rep}}$ is a short-range repulsive energy fitted to higher-level calculations.

We employed the 3ob-3-1 \cite{gaus2014parameterization} parameter set, which provides Slater-Koster integral tables and repulsive potentials optimized for organic and biological molecules containing C, H, N, O, S, and P atoms. This parameter set has been validated extensively for ground-state geometries and relative energetics of organic molecules. The self-consistent charge cycle was iterated until the maximum charge difference between iterations fell below $10^{-5}$ elementary charges, with a maximum of 250 iterations permitted before declaring convergence failure.

Geometry optimization was performed using the conjugate gradient algorithm with forces computed analytically from the DFTB energy expression. Optimization proceeded until the maximum force component on any atom fell below $10^{-4}$ Hartree per Bohr. Initial geometries were generated from SMILES representations using RDKit's distance geometry embedding followed by MMFF94 force field optimization \cite{tosco2014bringing}, providing reasonable starting structures for the quantum mechanical refinement.

From the converged self-consistent calculation, we extracted frontier orbital energies corresponding to the highest occupied molecular orbital (HOMO) and lowest unoccupied molecular orbital (LUMO). The HOMO energy $\varepsilon_{\text{HOMO}}$ approximates the negative of the ionization potential within Koopmans' theorem, while the LUMO energy $\varepsilon_{\text{LUMO}}$ approximates the negative of the electron affinity. The fundamental band gap was computed as the HOMO-LUMO difference:

$$E_g = \varepsilon_{\text{LUMO}} - \varepsilon_{\text{HOMO}}$$



\subsection*{Solvation Effects Validation}

We performed our DFTB  calculaiton on gas-phase due to its efficiency. Since the hydrogen evolution reaction occurs in aqueous media, we validated that design rules discovered from gas-phase calculations transfer to solvated conditions. We performed benchmark calculations on 50 representative molecules using DFTB+ with the Generalized Born Surface Area (GBSA) implicit solvation model for water.

The GBSA model treats the molecule as a continuous region with dielectric constant $\varepsilon_{\text{in}}$ surrounded by solvent with dielectric constant $\varepsilon_{\text{out}}$ = 80.2 (water). The solvation energy is computed through the Generalized Born equation with Onufriev-Bashford-Case (OBC) corrections for Born radii. We employed the GFN2-xTB/GBSA(water) parameterization including hydrogen bond corrections and solvent accessible surface area (SASA) terms.

The benchmark subset included all five champion molecules, 15 ether-containing molecules, 15 carbonyl-containing molecules, 10 halogenated molecules, and 5 baseline molecules. For each molecule, we compared frontier orbital energies between vacuum and solvated calculations.

Solvation effects were found to be minimal: the mean shifts were -0.016 ± 0.141 eV for HOMO, +0.013 ± 0.125 eV for LUMO, and +0.029 ± 0.118 eV for band gap. More importantly, vacuum and solvated calculations showed excellent correlation: $r$ = 0.9757 for HOMO, $r$ = 0.9890 for LUMO, and $r$ = 0.9907 for band gap. These correlations indicate that the relative ordering of molecules is preserved upon solvation, validating the use of gas-phase calculations for design rule discovery.

We further verified that the primary design rules hold in solvated calculations. The ether effect on HOMO energy remained significant with effect size $d$ = 1.94 ($p <$  0.0001). The carbonyl effect on band gap remained significant with $d$ = -1.04 ($p$ = 0.0006). The halogen effect on HOMO energy remained significant with $d$ = -1.03 ($p$ = 0.0008). In all cases, the effect sizes in solvated calculations were comparable to or larger than those observed in vacuum calculations, confirming that the discovered design rules are applicable to aqueous HER conditions.

\subsection*{Higher-Level Theory Validation}
To establish error bounds and confirm that design rules transfer to higher levels of theory, we performed benchmark calculations on 25 molecules using B3LYP/def2-SVP as implemented in PySCF \cite{sun2020recent}. This hybrid density functional provides a well-established reference for organic molecular properties and is widely used for photocatalyst calculations.

The benchmark subset included all five champion molecules plus 20 additional molecules spanning the functional group categories (ether, carbonyl, halogen, baseline), with smaller molecules prioritized for computational tractability. For each molecule, we compared DFTB+ predictions against B3LYP results.

DFTB+ showed excellent correlation with B3LYP for frontier orbital energies: $r$ = 0.9410 for HOMO (MAE = 0.73 eV, RMSE = 0.80 eV), $r$ = 0.9934 for LUMO (MAE = 0.71 eV, RMSE = 0.72 eV), and $r$ = 0.9274 for band gap (MAE = 1.44 eV, RMSE = 1.47 eV). DFTB+ systematically underestimates band gaps relative to B3LYP, which is expected given that tight-binding methods typically underestimate HOMO-LUMO gaps compared to hybrid functionals.

Critically, the design rules were preserved at the B3LYP level. Ether linkages elevated HOMO energies by 1.41 eV in the B3LYP calculations (compared to 0.58 eV in DFTB+). Carbonyl groups reduced band gaps (effect = -0.16 eV at B3LYP). Halogen substituents lowered HOMO energies by 0.58 eV at B3LYP (compared to 0.43 eV in DFTB+). The qualitative direction and relative magnitude of all effects were consistent across theory levels, validating that the discovered design rules represent genuine structure-property relationships rather than artifacts of the semi-empirical method.

\subsection*{Molecular Feature Extraction}
A critical methodological contribution of this work is the implementation of comprehensive, open-vocabulary feature extraction that goes beyond traditional hardcoded structural descriptors. Many computational discovery workflows suffer from the "streetlight effect": they can only discover correlations for features that were explicitly programmed into the analysis, potentially missing important structure-property relationships that the investigators did not anticipate. We addressed this limitation through a two-layer feature extraction system comprising 130 molecular descriptors.

The first layer employed all 85 functional group fragment descriptors available in the RDKit Fragments module. These descriptors, accessed programmatically through dynamic introspection of the module, cover a comprehensive range of organic functional groups including aldehydes, ketones, esters, amides, carboxylic acids, alcohols, ethers, epoxides, lactams, lactones, sulfones, sulfonamides, nitro groups, nitriles, azides, and many others. By using the complete RDKit fragment library rather than a manually curated subset, we ensured that the analysis could detect correlations with any functional group present in the dataset, not merely those we hypothesized to be important.

The second layer comprised 20 custom SMARTS patterns targeting structural motifs of particular relevance to photocatalyst design that are not fully captured by the RDKit fragment library. These patterns included heterocyclic systems such as benzothiadiazole, thiadiazole, oxadiazole, triazole, carbazole, phenothiazine, phenoxazine, and acridine. We also defined patterns for donor-acceptor architectures, extended conjugation motifs, and specific bonding patterns including ortho-disubstituted aromatics and spiro centers.

Additional descriptors extracted for each molecule included ring system properties such as total ring count, aromatic ring count, aliphatic ring count, heterocycle count, spiro atom count, and bridgehead atom count. The ring size distribution was characterized by counting rings of each size from 3-membered to 8-membered. Ring fusion patterns were quantified by counting atoms shared between multiple rings.

Electronic descriptors included counts of heteroatoms by element type (N, O, S, halogens), hydrogen bond donor and acceptor counts, aromatic atom count and fraction, and formal charge distribution. Topological descriptors included molecular weight, heavy atom count, rotatable bond count, topological polar surface area, Labute approximate surface area, and fraction of sp3-hybridized carbons.

In total, the expanded feature extraction system computed 130+ molecular descriptors for each molecule, representing a six-fold increase over the approximately 25 features used in conventional hardcoded analyses. This comprehensive feature space enabled discovery of structure-property relationships that would have been missed by a more limited analysis, as demonstrated by the identification of the halogenation effect described in the Results section.

\subsection*{Discovery Cycle Protocol}

\begin{figure}
    \centering
    \includegraphics[width=0.95\linewidth]{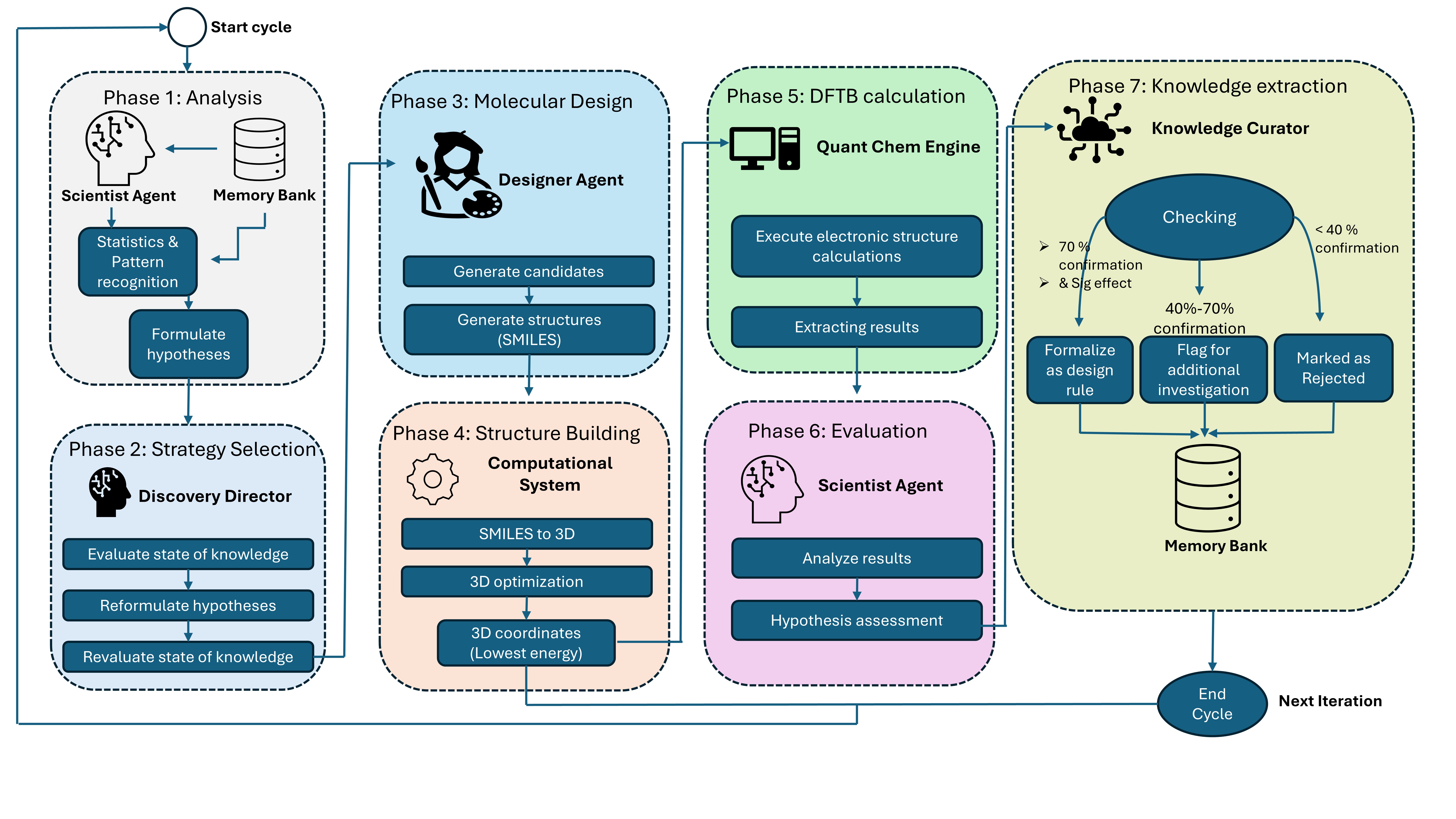}
    \caption{Schematic overview of the seven-phase Agentic Discovery Protocol. The workflow illustrates the iterative cycle starting from Analysis and Strategy Selection, moving through Molecular Design and Structure Building, and concluding with Quantum Calculation, Evaluation, and Knowledge Extraction. Specific agents (Scientist, Director, Designer, Curator) manage distinct decision points, utilizing statistical thresholds to confirm design rules or flag hypotheses for further investigation}
    \label{fig:protocol}
\end{figure}

Each discovery cycle followed a standardized protocol designed to ensure reproducibility and enable systematic comparison across cycles. The protocol, which is shown in Figure \ref{fig:protocol} comprised seven sequential phases: analysis, strategy selection, molecular design, structure building, quantum calculation, evaluation, and knowledge extraction.

During the analysis phase, the Scientist Agent queried the Memory Bank for all completed calculations and computed summary statistics stratified by the full set of 130 structural features. Pattern recognition employed both statistical hypothesis testing and qualitative reasoning about chemical mechanisms. For example, if molecules containing a particular functional group exhibited systematically different band gaps than molecules lacking that group, the agent would formulate a mechanistic hypothesis explaining the observation in terms of electronic effects such as induction, resonance, or hyperconjugation. The expanded feature set enabled the system to identify correlations that would have been invisible to a hardcoded analysis.

The strategy selection phase involved the Discovery Director evaluating the current state of knowledge and selecting which hypotheses to pursue in the current cycle. This selection balanced the confidence level of existing hypotheses against the potential information gain from testing uncertain or novel predictions. In early cycles, the system favored broad exploration to establish baseline understanding; in later cycles, it shifted toward focused validation of promising design rules.

During molecular design, the Designer Agent generated candidate structures to test selected hypotheses. For hypothesis validation experiments, this typically involved generating matched pairs of molecules: a set containing the structural feature under investigation and a control set lacking that feature but otherwise similar in scaffold and size. The agent generated 10 molecules per cycle, providing sufficient statistical power to detect medium-to-large effect sizes while maintaining computational tractability.

Structure building converted SMILES representations to three-dimensional atomic coordinates, followed by the procedure described in our previous publication \cite{peivaste2025hype}. The procedure employed RDKit's ETKDG (Experimental-Torsion Knowledge Distance Geometry) algorithm to generate initial conformers, followed by MMFF94 force field optimization to relieve steric clashes and establish reasonable bond lengths and angles. For molecules containing multiple rotatable bonds, three conformers were generated and the lowest-energy conformer after force field optimization was selected for quantum chemical calculation.

Quantum calculations proceeded as described in the Electronic Structure Calculations section. The evaluation phase compared calculation results to predictions derived from the hypothesis under test. The Scientist Agent computed effect sizes and statistical significance for the comparison between test and control groups, then assessed whether the results confirmed, refuted, or remained inconclusive regarding the hypothesis.

Knowledge extraction occurred when hypothesis confirmation rate exceeded 70\% with a statistically significant effect. The Knowledge Curator agent formalized the finding as a design rule, recording the structural feature, target property, direction and magnitude of effect, statistical evidence, and supporting molecule identifiers. Rules with confirmation rates below 40\% were marked as rejected; those between 40\% and 70\% were flagged for additional investigation in subsequent cycles.

Each discovery cycle completed in approximately 80 seconds on average (range: 44.9--125.3 seconds), enabling rapid iteration and real-time hypothesis refinement.

\subsection*{Targeted Validation Experiments}

Beyond the iterative discovery cycles, we conducted targeted validation experiments to rigorously test the most promising design rules identified during exploration. These validation experiments followed a pre-registered hypothesis testing framework with larger sample sizes than routine discovery cycles.

The carbonyl validation experiment tested the hypothesis that carbonyl-containing molecules exhibit reduced band gaps relative to non-carbonyl analogues. The Designer Agent generated molecules with explicit carbonyl groups (aldehydes, ketones, or quinones) and matched controls lacking carbonyl functionality but maintaining similar molecular weights and aromatic ring counts.

The ether validation experiment tested the hypothesis that ether linkages elevate HOMO energies. This experiment employed a two-phase design: an initial exploratory phase analyzing existing database molecules, followed by a confirmatory phase with newly generated molecules. The two-phase approach provides both discovery and validation components within a single experimental framework.

The cyano/conjugation validation experiment tested two related hypotheses: that cyano groups lower LUMO energies, and that extended conjugation narrows band gaps. Analysis stratified molecules by feature presence to assess each effect independently.

For all validation experiments, effect sizes were computed using Cohen's $d$, defined as the standardized mean difference between groups:

\begin{equation}
d = \frac{\bar{x}_1 - \bar{x}_2}{s_p}
\end{equation}

where $\bar{x}_1$ and $\bar{x}_2$ are the sample means of the two groups, and $s_p$ is the pooled standard deviation:

\begin{equation}
s_p = \sqrt{\frac{(n_1 - 1)s_1^2 + (n_2 - 1)s_2^2}{n_1 + n_2 - 2}}
\end{equation}

Here $n_1$ and $n_2$ denote sample sizes and $s_1$ and $s_2$ denote sample standard deviations. Effect sizes were interpreted following conventional thresholds: $|d| < 0.5$ as small, $0.5 \leq |d| < 0.8$ as medium, and $|d| \geq 0.8$ as large.

\subsection*{Statistical Analysis}

All statistical analyses were performed using Python 3.11 with NumPy 1.24 for numerical computation, SciPy 1.11 for statistical testing, and Pandas 2.0 for data manipulation. Statistical significance was assessed using Welch's $t$-test \cite{welch1947generalization} for independent samples with unequal variances, implemented as:

\begin{equation}
t = \frac{\bar{x}_1 - \bar{x}_2}{\sqrt{\frac{s_1^2}{n_1} + \frac{s_2^2}{n_2}}}
\end{equation}

where the degrees of freedom were computed using the Welch-Satterthwaite \cite{satterthwaite1946approximate} approximation:

\begin{equation}
\nu = \frac{\left(\frac{s_1^2}{n_1} + \frac{s_2^2}{n_2}\right)^2}{\frac{s_1^4}{n_1^2(n_1-1)} + \frac{s_2^4}{n_2^2(n_2-1)}}
\end{equation}

The Welch test was chosen over the standard Student's $t$-test because preliminary analysis indicated heterogeneous variances across structural groups, violating the equal-variance assumption of the standard test.

For the multi-feature comparative analysis, we computed pairwise comparisons between molecules with and without each structural feature of interest across all extracted features. To control the family-wise error rate across multiple comparisons, we applied the Holm-Bonferroni correction \cite{holm1979simple}, which orders $p$-values and compares each to a sequentially adjusted significance threshold. A feature was considered statistically significant if its corrected $p$-value fell below 0.05. The comprehensive feature analysis identified 32 features with significant correlations to at least one electronic property, from which the six primary design rules were extracted based on effect size magnitude and mechanistic interpretability.

Confidence intervals for Cohen's $d$ were computed using the noncentral $t$-distribution approximation \cite{hedges1981distribution}. The 95\% confidence interval provides a range of plausible effect sizes consistent with the observed data, enabling assessment of both statistical significance (whether the interval excludes zero) and practical significance (whether the interval falls entirely within a range of substantively meaningful effects).

Feature interaction analysis examined whether combining multiple design strategies produced additive, synergistic, or antagonistic effects. We computed the expected band gap under an additive model by summing individual feature effects, then compared this prediction to the observed band gap for molecules containing both features. The interaction term $\Delta_{\text{int}}$ was computed as:

\begin{equation}
\Delta_{\text{int}} = E_g^{\text{both}} - \left( E_g^{\text{neither}} - \Delta_{\text{feature1}} - \Delta_{\text{feature2}} \right)
\end{equation}

where $E_g^{\text{both}}$ is the mean band gap for molecules with both features, $E_g^{\text{neither}}$ is the mean for molecules with neither feature, and $\Delta_{\text{feature1}}$ and $\Delta_{\text{feature2}}$ are the individual feature effects (mean difference from the neither group). Negative interaction terms indicate diminishing returns when combining strategies; positive terms indicate synergistic enhancement.

\subsection*{Implementation Details}

The ChemNavigator AI system was implemented in Python using a modular architecture that separates agent logic from infrastructure components. LLM capabilities were accessed through the Google API using Gemini 2.5 Pro (model ID: gemini-2.5-pro) for agent reasoning tasks. Prompts for each agent were developed through iterative refinement to ensure consistent output formatting and reliable extraction of structured information from natural language responses.

The Memory Bank was implemented using SQLite as the storage backend, with tables for molecules, calculations, hypotheses, and design rules. Database queries were constructed using parameterized SQL to prevent injection vulnerabilities and ensure reproducibility. The schema supported efficient filtering by molecular properties, structural features, and temporal metadata.

Molecular manipulation relied on RDKit 2023.09 for SMILES parsing, substructure searching, and descriptor calculation. The expanded feature extraction module employed dynamic introspection of the RDKit Fragments module to automatically discover and compute all available fragment descriptors, ensuring that the analysis benefited from the complete RDKit functional group vocabulary without manual enumeration. Three-dimensional coordinate generation used the ETKDG algorithm with random seed 42 for reproducibility. The Atomic Simulation Environment (ASE) 3.22 \cite{larsen2017atomic} provided utilities for coordinate transformation and file format conversion.

DFTB+ calculations were executed through a Python wrapper that generated Human-Readable Structured Data (HSD) input files, invoked the DFTB+ executable, and parsed output files for orbital energies and total energies. The wrapper implemented timeout handling (300 seconds per calculation) and automatic retry with adjusted convergence parameters for calculations that failed on the first attempt. Higher-level DFT calculations for benchmarking used PySCF with the def2-SVP basis set.

All experiments were conducted on a workstation equipped with an AMD Ryzen 9 5900X processor (12 cores, 24 threads) and 64 GB RAM. Each discovery cycle completed in approximately 80 seconds on average (range: 44.9--125.3 seconds), encompassing hypothesis generation, molecular design, geometry optimization, electronic structure calculation, and statistical analysis for 10 molecules. The complete discovery campaign comprising 20 cycles across four runs required approximately 27 minutes of total computation time. This rapid iteration—approximately 8 seconds per molecule from SMILES generation through property calculation—enables real-time hypothesis testing and design rule refinement that would be impractical with higher-level quantum chemical methods.

Source code, molecular datasets, calculation outputs, and analysis scripts are available at at \href{https://github.com/Iman-Peivaste/ChemNavigator}{github.com/Iman-Peivaste/ChemNavigator}.

\subsection*{Reproducibility Considerations}

Several measures were implemented to ensure reproducibility. Random seeds were fixed for all stochastic operations including conformer generation (seed 42). The Memory Bank maintained complete provenance records linking each molecule to its source and each calculation to its input parameters.

The autonomous discovery process introduces inherent variability through the language model's reasoning. To characterize this variability, we observed consistent identification of the major design rules across the four independent runs, suggesting that the discovered rules reflect robust patterns in the underlying data rather than artifacts of any single discovery trajectory.

The comprehensive feature extraction approach provides an additional layer of reproducibility assurance. Because the system analyzes all available features rather than a manually selected subset, the analysis is less susceptible to investigator bias in feature selection.

All statistical analyses were performed on the complete dataset without exclusion of outliers. The reported effect sizes and $p$-values reflect the full variability present in the calculated molecular properties.

\section*{Data Availability} \label{data_av}

The seed data used in this study originates from the source described in \cite{li2021combining}

\section*{Code Availability} \label{code_av}

The Python framework is accessible at \href{https://github.com/Iman-Peivaste/ChemNavigator}{github.com/Iman-Peivaste/ChemNavigator}.

\section*{Acknowledgements}

This work was funded by the Luxembourg Fonds National de la Recherche (FNR) through the grant PRIDE21/16758661/HYMAT.  

\section*{Author Contributions}

Iman Peivaste: Developing the initial concept and workflow, framework design, writing the original draft, visualization, methodology, investigation, and formal analysis. 

Ahmed Makradi: Writing, review, editing, resources, methodology, and formal analysis.

Salim Belouettar: Developing the initial concept and workflow, Writing, review, editing, supervision, methodology, funding acquisition, data curation, and conceptualization

\section*{Competing Interests}

The authors declare that they have no known competing financial interests or personal relationships that could have appeared to influence the work reported in this paper.

\clearpage
\bibliographystyle{naturemag}. 
\bibliography{refs}

\end{document}